\documentclass[a4paper]{article}
\usepackage[ansinew]{inputenc}
\usepackage{times}
\usepackage[T1]{fontenc}
\usepackage{graphicx}
\usepackage{geometry}
\usepackage{amssymb}
\usepackage{amsmath}

\usepackage{rotating}

\usepackage{xcolor}
\colorlet{shadecolor}{gray!25}
\usepackage{subfig}
\usepackage{float}

\author{\normalsize Ludwig A. Hothorn,\\ 
\footnotesize Im Grund 12, D-31867 Lauenau, Germany (e-mail:ludwig@hothorn.de)\ \scriptsize(retired from Leibniz University Hannover)}

\title{Simultaneous inference of correlated marginal tests using intersection-union or union-intersection test principle}
\begin{document}

\maketitle
\begin{abstract}
Two main approaches in simultaneous inference are intersection-union tests and union-intersection tests. For intersection-union hypotheses, the classical IUT based on marginal p-values and the all-in-alternative UIT are compared. Depending on correlation, number of marginal tests and patterns of the alternative the inherent power loss of the aiaUIT seems to be acceptable, considering its advantage, namely the availability of simple-to-interpret simultaneous confidence interval.
\end{abstract}

\section{Introduction}\label{sec1}
Although the majority of multiple testing is based on the union-intersection test (UIT) principle, some applications exist which use the complementary intersection-union test (IUT) principle, e.g. the analysis of multiple correlated primary endpoints \cite{Hothorn2003b}, \cite{Quan2001}, or fixed dose combinations \cite{Sidik2003}.\\

Simultaneous inference considering $j$ elementary hypotheses of interest with related correlated test statistics $T_j$ and compatible p-value $p_j$ based therefore on two main principles (under some conditions): intersection-union tests (IUT) and union-intersection tests (UIT). \\
Their global $H_0$'s are:\\
UIT: $H_0=\bigcap_{j\in J}H_{0,j}$ and $T^{max}=max(T_1,...,T_j, ...,T_J)$ or $p^{min}=min(p_1,...,p_j,...,p_J)$\\
IUT: $H_0=\bigcup_{j\in J}H_{0,j}$ and $T^{min}=min(T_1,...,T_j, ...,T_J)$ or $p^{max}=max(p_1,...,p_j,...,p_J)$\\
Assuming uncorrelated tests, their behavior under $H_0$ seems to be related:\\
UIT: $H_A: \exists^{(any j)} p^{j}<\alpha/J)$\\
IUT: $H_A:\forall (p_1,...,p_j,...,p_J)<\alpha$\\
whereas considering the more general and more realistic case of correlated tests:\\
UIT: $H_A: \exists^{(any j)} T^{j}<t_{J, df, R, \alpha}$\\
IUT: $H_A:\forall (p_1,...,p_j,...,p_J)<\alpha$\\

I.e. the IUT can be used for the correlated case as well. A related solution based on the multivariate t distribution exists \cite{Hasler2009}, but the consequence is that the joint test rejects global $H_0$ even when some elementary tests remain still under  $H_0$ which seem to be hard to interpret in most applications.\\
The second argument is that UIT based on $\alpha$-adjustment, where IUT does not. The second argument is a bit superficial, since both tests are conservative as dimension J increases, the IUT just about the 'and... and' condition. An essential difference is that due to the Gabriel theorem  \cite{Gabriel1969} the UIT makes individual decisions available, the IUT however provide only the global decision. This is an essential disadvantage for some applications.
An alternative is the use of the \textit{all-in-the-alternative UITs} (aiaUIT) instead of classical IUT tests discussed in the following. Related is the concept of all-pairs power \cite{Ramsey1978} or complete power UIT approach \cite{Westfall2011} \cite{Liu2011}. \\
Looking more closely at $H_0$ formulation, the UIT under $H_1$ is formulated for all possible patterns of individual decisions: from exactly one test under $H^A$ to exactly all tests or all partitions in between. Assuming $J=4$ for example, the following elementary outcomes are possible under UIT's $H_A$:\\
\footnotesize
$[H_A^1, H_0^2, H_0^3, H_0^4]$\\
$[H_{01}, H_{A2}, H_{03}, H_{04}]$\\
$[H_{01}, H_{02}, H_{A3}, H_{04}]$\\
$[H_{01}, H_{02}, H_{03}, H_{A4}]$\\
$[H_{A1}, H_{A2}, H_{03}, H_{04}]$\\
$[H_{A1}, H_{02}, H_{A3}, H_{04}]$\\
$[H_{A1}, H_{A2}, H_{A3}, H_{04}]$\\
$[H_{A1}, H_{A2}, H_{03}, H_{A4}]$\\
$[H_{01}, H_{A2}, H_{A3}, H_{04}]$\\
$[H_{01}, H_{A2}, H_{03}, H_{A4}]$\\
$[H_{01}, H_{02}, H_{A3}, H_{04}]$\\
$[H_{A1}, H_{A2}, H_{A3}, H_{A4}]$\\
\normalsize
The last set reveals the same conclusion like an IUT, but it provides individual outcomes, such as simultaneous confidence interval or adjusted p-values. This property can be used applying an UIT in an IUT hypothesis scenario. The question is, how high is the inherent power loss? This is answered by a simulation study with different scenarios in the following section considering the correlation between the test as a primary criteria, the pattern of individual $H^A$'s and the number and kind of tests.

\section{Simulation studies}
The primary interest is the comparison of classical IUT with the aiaIUT under $H_0$ and patterns of $H^A$ by means of the power ratio (RR) (as  as a rough empirical measure of a Pitman efficiency)
\subsection{IUT for two correlated primary endpoints}
In phase II-III randomized clinical trials sometime two primary endpoints occur. The benefit of the drug is demonstrated, exactly if both endpoints are under the alternative, using the classic IUT as $max(p_j)$-test.

\begin{table}[ht]
\centering
\scalebox{0.5}{
\begin{tabular}{rrrrrrrrrr|rr|rr|rr|r|r}
  \hline
 & n1 & n2 & ma1 & ma2 & mb1 & mb2 & sa & sb & rho & IUT & UIT & m1 & m2 & e1 & e2 & aiaIUT & RR \\ 
  \hline
H0 & 20 & 20.0 & 1.0 & 1.0 & 10.0 & 10.0 & 1.0 & 11.0 & 0.900 & 0.031 & 0.044 & 0.031 & 0.041 & 0.046 & 0.052 & 0.028 & 0.903 \\ 
   \hline
1 & 20 & 20.0 & 1.0 & 1.0 & 10.0 & 18.0 & 1.0 & 11.0 & 0.900 & 0.046 & 0.682 & 0.031 & 0.682 & 0.046 & 0.735 & 0.031 & 0.674 \\ 
1 & 20 & 20.0 & 1.0 & 1.1 & 10.0 & 18.0 & 1.0 & 11.0 & 0.900 & 0.077 & 0.682 & 0.057 & 0.682 & 0.077 & 0.735 & 0.057 & 0.740 \\ 
1 & 20 & 20.0 & 1.0 & 1.2 & 10.0 & 18.0 & 1.0 & 11.0 & 0.900 & 0.148 & 0.682 & 0.110 & 0.682 & 0.148 & 0.735 & 0.110 & 0.743 \\ 
1 & 20 & 20.0 & 1.0 & 1.3 & 10.0 & 18.0 & 1.0 & 11.0 & 0.900 & 0.233 & 0.682 & 0.194 & 0.682 & 0.233 & 0.735 & 0.194 & 0.833 \\ 
1 & 20 & 20.0 & 1.0 & 1.4 & 10.0 & 18.0 & 1.0 & 11.0 & 0.900 & 0.331 & 0.684 & 0.280 & 0.682 & 0.331 & 0.735 & 0.278 & 0.840 \\ 
1 & 20 & 20.0 & 1.0 & 1.5 & 10.0 & 18.0 & 1.0 & 11.0 & 0.900 & 0.459 & 0.687 & 0.391 & 0.682 & 0.463 & 0.735 & 0.386 & 0.841 \\ 
1 & 20 & 20.0 & 1.0 & 1.6 & 10.0 & 18.0 & 1.0 & 11.0 & 0.900 & 0.571 & 0.700 & 0.526 & 0.682 & 0.587 & 0.735 & 0.508 & 0.890 \\ 
1 & 20 & 20.0 & 1.0 & 1.7 & 10.0 & 18.0 & 1.0 & 11.0 & 0.900 & 0.654 & 0.731 & 0.648 & 0.682 & 0.699 & 0.735 & 0.599 & 0.916 \\ 
\hline
1 & 20 & 20.0 & 1.0 & 1.0 & 10.0 & 10.0 & 1.0 & 11.0 & 0.700 & 0.019 & 0.050 & 0.028 & 0.029 & 0.046 & 0.054 & 0.007 & 0.368 \\ 
1 & 20 & 20.0 & 1.0 & 1.0 & 10.0 & 18.0 & 1.0 & 11.0 & 0.700 & 0.046 & 0.671 & 0.028 & 0.671 & 0.046 & 0.749 & 0.028 & 0.609 \\ 
1 & 20 & 20.0 & 1.0 & 1.1 & 10.0 & 18.0 & 1.0 & 11.0 & 0.700 & 0.076 & 0.672 & 0.048 & 0.671 & 0.077 & 0.749 & 0.047 & 0.618 \\ 
1 & 20 & 20.0 & 1.0 & 1.2 & 10.0 & 18.0 & 1.0 & 11.0 & 0.700 & 0.146 & 0.673 & 0.092 & 0.671 & 0.148 & 0.749 & 0.090 & 0.616 \\ 
1 & 20 & 20.0 & 1.0 & 1.3 & 10.0 & 18.0 & 1.0 & 11.0 & 0.700 & 0.227 & 0.674 & 0.162 & 0.671 & 0.233 & 0.749 & 0.159 & 0.700 \\ 
1 & 20 & 20.0 & 1.0 & 1.4 & 10.0 & 18.0 & 1.0 & 11.0 & 0.700 & 0.315 & 0.682 & 0.258 & 0.671 & 0.331 & 0.749 & 0.247 & 0.784 \\ 
1 & 20 & 20.0 & 1.0 & 1.5 & 10.0 & 18.0 & 1.0 & 11.0 & 0.700 & 0.432 & 0.696 & 0.356 & 0.671 & 0.463 & 0.749 & 0.331 & 0.766 \\ 
1 & 20 & 20.0 & 1.0 & 1.6 & 10.0 & 18.0 & 1.0 & 11.0 & 0.700 & 0.536 & 0.723 & 0.494 & 0.671 & 0.587 & 0.749 & 0.442 & 0.825 \\ 
1 & 20 & 20.0 & 1.0 & 1.7 & 10.0 & 18.0 & 1.0 & 11.0 & 0.700 & 0.616 & 0.759 & 0.611 & 0.671 & 0.699 & 0.749 & 0.523 & 0.849 \\
\hline 
1 & 20 & 20.0 & 1.0 & 1.0 & 10.0 & 10.0 & 1.0 & 11.0 & 0.090 & 0.001 & 0.057 & 0.024 & 0.034 & 0.046 & 0.053 & 0.001 & 1.000 \\ 
1 & 20 & 20.0 & 1.0 & 1.0 & 10.0 & 18.0 & 1.0 & 11.0 & 0.090 & 0.041 & 0.652 & 0.024 & 0.648 & 0.046 & 0.748 & 0.020 & 0.488 \\ 
1 & 20 & 20.0 & 1.0 & 1.1 & 10.0 & 18.0 & 1.0 & 11.0 & 0.090 & 0.068 & 0.656 & 0.044 & 0.648 & 0.077 & 0.748 & 0.036 & 0.529 \\ 
1 & 20 & 20.0 & 1.0 & 1.2 & 10.0 & 18.0 & 1.0 & 11.0 & 0.090 & 0.129 & 0.662 & 0.082 & 0.648 & 0.148 & 0.748 & 0.068 & 0.527 \\ 
1 & 20 & 20.0 & 1.0 & 1.3 & 10.0 & 18.0 & 1.0 & 11.0 & 0.090 & 0.191 & 0.680 & 0.148 & 0.648 & 0.233 & 0.748 & 0.116 & 0.607 \\ 
1 & 20 & 20.0 & 1.0 & 1.4 & 10.0 & 18.0 & 1.0 & 11.0 & 0.090 & 0.255 & 0.712 & 0.230 & 0.648 & 0.331 & 0.748 & 0.166 & 0.651 \\ 
1 & 20 & 20.0 & 1.0 & 1.5 & 10.0 & 18.0 & 1.0 & 11.0 & 0.090 & 0.359 & 0.756 & 0.327 & 0.648 & 0.463 & 0.748 & 0.219 & 0.610 \\ 
1 & 20 & 20.0 & 1.0 & 1.6 & 10.0 & 18.0 & 1.0 & 11.0 & 0.090 & 0.455 & 0.799 & 0.463 & 0.648 & 0.587 & 0.748 & 0.312 & 0.686 \\ 
1 & 20 & 20.0 & 1.0 & 1.7 & 10.0 & 18.0 & 1.0 & 11.0 & 0.090 & 0.535 & 0.838 & 0.581 & 0.648 & 0.699 & 0.748 & 0.391 & 0.731 \\ 
   \hline
\end{tabular}
}
\caption{Simulated power for 2 correlated endpoints in 2-sample design}
\end{table}
\normalsize

The power loss, represented by RR, becomes smaller, the  higher the correlation between the tests and the more similar the elementary powers (e1,e2) are.

%%%%%%%%%%%%%%%%%%%%%%%%%%%%%%%%%%%%%%%%%%%%%%%%%%%%%%%%%%%%%%%%%%%%%%%%%%%%%%%%%%%%%%%%
\subsection{IUT for three correlated primary endpoints}
The question whether the loss of power decreases with increasing number of elementary tests (due to the considerable conservatism of the classical IUT) is answered on the one hand with 3 and 4 endpoints in the two-sample design and on the other hand for 3 endpoints and two simultaneous treatment contrasts. For the global UIT and aiaUIT the multiple marginal model approach \cite{Schaarschmidt2021} was used for this multiple endpoints scenario \cite{Hothorn2021aa} (see the R-code in the Appendix).

\begin{table}[ht]
\centering
\scalebox{0.5}{
\begin{tabular}{rrrrrrrrrrrrrrr|r|r|rrr|rrr|r||r}
  \hline
 & n1 & n2 & ma1 & ma2 & mb1 & mb2 & mc1 & mc2 & sa & sb & sc & rho1 & rho2 & rho3 & IUT & UIT & m1 & m2 & m3 & e1 & e2 & e3 & aiaUIT & RR \\ 
  \hline
H0 & 20 & 20 & 1.0 & 1.0 & 2.0 & 1.0 & 10.0 & 10.0 & 1.0 & 4.0 & 11.00 & 0.90 & 0.90 & 0.90 & 0.009 & 0.042 & 0.030 & 0.006 & 0.033 & 0.051 & 0.010 & 0.051 & 0.005 & 0.556 \\ 
   \hline 
1 & 20 & 20 & 1.0 & 1.0 & 2.0 & 1.0 & 10.0 & 10.0 & 1.0 & 4.0 & 11.00 & 0.09 & 0.09 & 0.09 & 0.000 & 0.026 & 0.011 & 0.005 & 0.011 & 0.045 & 0.007 & 0.044 & 0.000 &  \\ 
   
1 & 20 & 20 & 1.0 & 1.0 & 2.0 & 5.0 & 10.0 & 18.0 & 1.0 & 4.0 & 11.00 & 0.90 & 0.90 & 0.90 & 0.056 & 0.722 & 0.029 & 0.658 & 0.646 & 0.056 & 0.753 & 0.730 & 0.029 & 0.518 \\ 

1 & 20 & 20 & 1.0 & 1.1 & 2.0 & 5.0 & 10.0 & 18.0 & 1.0 & 4.0 & 11.00 & 0.90 & 0.90 & 0.90 & 0.082 & 0.716 & 0.053 & 0.649 & 0.626 & 0.082 & 0.734 & 0.717 & 0.053 & 0.646 \\ 
  
1 & 20 & 20 & 1.0 & 1.2 & 2.0 & 5.0 & 10.0 & 18.0 & 1.0 & 4.0 & 11.00 & 0.90 & 0.90 & 0.90 & 0.151 & 0.723 & 0.103 & 0.672 & 0.628 & 0.151 & 0.757 & 0.729 & 0.103 & 0.682 \\ 

1 & 20 & 20 & 1.0 & 1.3 & 2.0 & 5.0 & 10.0 & 18.0 & 1.0 & 4.0 & 11.00 & 0.90 & 0.90 & 0.90 & 0.246 & 0.742 & 0.177 & 0.684 & 0.666 & 0.246 & 0.757 & 0.750 & 0.177 & 0.720 \\ 
1 & 20 & 20 & 1.0 & 1.4 & 2.0 & 5.0 & 10.0 & 18.0 & 1.0 & 4.0 & 11.00 & 0.90 & 0.90 & 0.90 & 0.321 & 0.711 & 0.258 & 0.652 & 0.628 & 0.322 & 0.733 & 0.715 & 0.257 & 0.801 \\ 
1 & 20 & 20 & 1.0 & 1.5 & 2.0 & 5.0 & 10.0 & 18.0 & 1.0 & 4.0 & 11.00 & 0.90 & 0.90 & 0.90 & 0.451 & 0.732 & 0.371 & 0.675 & 0.651 & 0.459 & 0.757 & 0.747 & 0.365 & 0.809 \\ 
1 & 20 & 20 & 1.0 & 1.6 & 2.0 & 5.0 & 10.0 & 18.0 & 1.0 & 4.0 & 11.00 & 0.90 & 0.90 & 0.90 & 0.570 & 0.751 & 0.501 & 0.695 & 0.663 & 0.598 & 0.766 & 0.746 & 0.477 & 0.837 \\ 
1 & 20 & 20 & 1.0 & 1.7 & 2.0 & 5.0 & 10.0 & 18.0 & 1.0 & 4.0 & 11.00 & 0.90 & 0.90 & 0.90 & 0.612 & 0.739 & 0.607 & 0.667 & 0.634 & 0.682 & 0.743 & 0.715 & 0.534 & 0.873 \\ 
1 & 20 & 20 & 1.0 & 1.8 & 2.0 & 5.0 & 10.0 & 18.0 & 1.0 & 4.0 & 11.00 & 0.90 & 0.90 & 0.90 & 0.671 & 0.768 & 0.726 & 0.666 & 0.642 & 0.796 & 0.749 & 0.734 & 0.577 & 0.860 \\ 
   \hline
1 & 20 & 20 & 1.0 & 1.0 & 2.0 & 5.0 & 10.0 & 18.0 & 1.0 & 4.0 & 11.00 & 0.90 & 0.90 & 0.90 & 0.056 & 0.722 & 0.029 & 0.658 & 0.646 & 0.056 & 0.753 & 0.730 & 0.029 & 0.518 \\ 
   \hline 
	
1 & 20 & 20 & 1.0 & 1.0 & 2.0 & 5.0 & 10.0 & 18.0 & 1.0 & 4.0 & 11.00 & 0.09 & 0.09 & 0.09 & 0.036 & 0.801 & 0.011 & 0.570 & 0.569 & 0.045 & 0.759 & 0.749 & 0.007 & 0.194 \\ 
1 & 20 & 20 & 1.0 & 1.1 & 2.0 & 5.0 & 10.0 & 18.0 & 1.0 & 4.0 & 11.00 & 0.09 & 0.09 & 0.09 & 0.061 & 0.805 & 0.026 & 0.570 & 0.569 & 0.087 & 0.759 & 0.749 & 0.013 & 0.213 \\ 
1 & 20 & 20 & 1.0 & 1.2 & 2.0 & 5.0 & 10.0 & 18.0 & 1.0 & 4.0 & 11.00 & 0.09 & 0.09 & 0.09 & 0.096 & 0.814 & 0.064 & 0.570 & 0.569 & 0.145 & 0.759 & 0.749 & 0.029 & 0.302 \\ 
1 & 20 & 20 & 1.0 & 1.3 & 2.0 & 5.0 & 10.0 & 18.0 & 1.0 & 4.0 & 11.00 & 0.09 & 0.09 & 0.09 & 0.143 & 0.823 & 0.110 & 0.570 & 0.569 & 0.227 & 0.759 & 0.749 & 0.048 & 0.336 \\ 
1 & 20 & 20 & 1.0 & 1.4 & 2.0 & 5.0 & 10.0 & 18.0 & 1.0 & 4.0 & 11.00 & 0.09 & 0.09 & 0.09 & 0.215 & 0.835 & 0.179 & 0.570 & 0.569 & 0.340 & 0.759 & 0.749 & 0.076 & 0.353 \\ 
1 & 20 & 20 & 1.0 & 1.5 & 2.0 & 5.0 & 10.0 & 18.0 & 1.0 & 4.0 & 11.00 & 0.09 & 0.09 & 0.09 & 0.287 & 0.853 & 0.277 & 0.570 & 0.569 & 0.462 & 0.759 & 0.749 & 0.110 & 0.383 \\ 
1 & 20 & 20 & 1.0 & 1.6 & 2.0 & 5.0 & 10.0 & 18.0 & 1.0 & 4.0 & 11.00 & 0.09 & 0.09 & 0.09 & 0.357 & 0.874 & 0.386 & 0.570 & 0.569 & 0.576 & 0.759 & 0.749 & 0.145 & 0.406 \\ 
1 & 20 & 20 & 1.0 & 1.7 & 2.0 & 5.0 & 10.0 & 18.0 & 1.0 & 4.0 & 11.00 & 0.09 & 0.09 & 0.09 & 0.413 & 0.890 & 0.507 & 0.570 & 0.569 & 0.687 & 0.759 & 0.749 & 0.186 & 0.450 \\ 
1 & 20 & 20 & 1.0 & 1.8 & 2.0 & 5.0 & 10.0 & 18.0 & 1.0 & 4.0 & 11.00 & 0.09 & 0.09 & 0.09 & 0.478 & 0.913 & 0.623 & 0.570 & 0.569 & 0.795 & 0.759 & 0.749 & 0.228 & 0.477 \\ 
   \hline
	1 & 20 & 20 & 1.0 & 1.8 & 2.0 & 5.0 & 10.0 & 18.0 & 1.0 & 4.0 & 11.00 & 0.09 & 0.09 & 0.09 & 0.478 & 0.913 & 0.623 & 0.570 & 0.569 & 0.795 & 0.759 & 0.749 & 0.228 & 0.477 \\ 
\end{tabular}
}
\caption{Simulated power for 3 correlated endpoints in 2-sample design}
\end{table}

\subsection{IUT for four correlated endpoints}

\begin{table}[ht]
\centering
\scalebox{0.5}{
\begin{tabular}{rrrrrrrrrrrr|r|r|rrrrrr|rrrrrr|r|r}
  \hline
 & n1 & n2 & ma1 & ma2 & mb1 & mb2 & mc1 & mc2 & md1 & md2 & sa & sb & sc & sd & rho & IUT & UIT & m1 & m2 & m3 & m4 & e1 & e2 & e3 & e4 & aiaUIT & RR \\ 
  \hline
1 & 20 & 20 & 1.0 & 1.0 & 2.0 & 2.0 & 10.0 & 10.0 & 0.1 & 0.1 & 1.0 & 4.00 & 11.00 & 0.50 & 0.90 & 0.021 & 0.045 & 0.027 & 0.028 & 0.033 & 0.033 & 0.052 & 0.042 & 0.057 & 0.048 & 0.015 & 0.714 \\ 
1 & 20 & 20 & 1.0 & 1.0 & 2.0 & 5.0 & 10.0 & 18.0 & 0.1 & 0.5 & 1.0 & 4.00 & 11.00 & 0.50 & 0.90 & 0.061 & 0.723 & 0.039 & 0.668 & 0.639 & 0.639 & 0.061 & 0.752 & 0.727 & 0.788 & 0.039 & 0.639 \\ 
1 & 20 & 20 & 1.0 & 1.1 & 2.0 & 5.0 & 10.0 & 18.0 & 0.1 & 0.5 & 1.0 & 4.00 & 11.00 & 0.50 & 0.90 & 0.102 & 0.723 & 0.067 & 0.668 & 0.637 & 0.637 & 0.102 & 0.752 & 0.728 & 0.788 & 0.067 & 0.657 \\ 
1 & 20 & 20 & 1.0 & 1.2 & 2.0 & 5.0 & 10.0 & 18.0 & 0.1 & 0.5 & 1.0 & 4.00 & 11.00 & 0.50 & 0.90 & 0.154 & 0.723 & 0.119 & 0.668 & 0.637 & 0.637 & 0.154 & 0.752 & 0.728 & 0.788 & 0.119 & 0.773 \\ 
1 & 20 & 20 & 1.0 & 1.3 & 2.0 & 5.0 & 10.0 & 18.0 & 0.1 & 0.5 & 1.0 & 4.00 & 11.00 & 0.50 & 0.90 & 0.241 & 0.732 & 0.178 & 0.675 & 0.644 & 0.644 & 0.241 & 0.755 & 0.730 & 0.791 & 0.178 & 0.739 \\ 
1 & 20 & 20 & 1.0 & 1.4 & 2.0 & 5.0 & 10.0 & 18.0 & 0.1 & 0.5 & 1.0 & 4.00 & 11.00 & 0.50 & 0.90 & 0.350 & 0.737 & 0.283 & 0.675 & 0.660 & 0.661 & 0.354 & 0.751 & 0.744 & 0.797 & 0.280 & 0.800 \\ 
1 & 20 & 20 & 1.0 & 1.5 & 2.0 & 5.0 & 10.0 & 18.0 & 0.1 & 0.5 & 1.0 & 4.00 & 11.00 & 0.50 & 0.90 & 0.458 & 0.744 & 0.378 & 0.691 & 0.661 & 0.662 & 0.468 & 0.764 & 0.741 & 0.801 & 0.371 & 0.810 \\ 
1 & 20 & 20 & 1.0 & 1.6 & 2.0 & 5.0 & 10.0 & 18.0 & 0.1 & 0.5 & 1.0 & 4.00 & 11.00 & 0.50 & 0.90 & 0.569 & 0.740 & 0.493 & 0.685 & 0.656 & 0.656 & 0.598 & 0.758 & 0.739 & 0.802 & 0.467 & 0.821 \\ 
1 & 20 & 20 & 1.0 & 1.7 & 2.0 & 5.0 & 10.0 & 18.0 & 0.1 & 0.5 & 1.0 & 4.00 & 11.00 & 0.50 & 0.90 & 0.640 & 0.759 & 0.618 & 0.692 & 0.674 & 0.673 & 0.701 & 0.757 & 0.749 & 0.814 & 0.557 & 0.870 \\ 
1 & 20 & 20 & 1.0 & 1.8 & 2.0 & 5.0 & 10.0 & 18.0 & 0.1 & 0.5 & 1.0 & 4.00 & 11.00 & 0.50 & 0.90 & 0.659 & 0.793 & 0.734 & 0.685 & 0.664 & 0.663 & 0.803 & 0.757 & 0.739 & 0.800 & 0.597 & 0.906 \\ 

  \hline   
1 & 20 & 20 & 1.0 & 1.0 & 2.0 & 2.0 & 10.0 & 10.0 & 0.1 & 0.1 & 1.0 & 4.00 & 11.00 & 0.50 & 0.09 & 0.000 & 0.041 & 0.010 & 0.017 & 0.016 & 0.016 & 0.047 & 0.043 & 0.049 & 0.034 & 0.000 &  \\ 
1 & 20 & 20 & 1.0 & 1.0 & 2.0 & 5.0 & 10.0 & 18.0 & 0.1 & 0.5 & 1.0 & 4.00 & 11.00 & 0.50 & 0.09 & 0.026 & 0.795 & 0.010 & 0.579 & 0.546 & 0.546 & 0.047 & 0.757 & 0.738 & 0.798 & 0.003 & 0.115 \\ 
1 & 20 & 20 & 1.0 & 1.1 & 2.0 & 5.0 & 10.0 & 18.0 & 0.1 & 0.5 & 1.0 & 4.00 & 11.00 & 0.50 & 0.09 & 0.051 & 0.796 & 0.032 & 0.579 & 0.546 & 0.546 & 0.083 & 0.757 & 0.738 & 0.798 & 0.012 & 0.235 \\ 
1 & 20 & 20 & 1.0 & 1.2 & 2.0 & 5.0 & 10.0 & 18.0 & 0.1 & 0.5 & 1.0 & 4.00 & 11.00 & 0.50 & 0.09 & 0.082 & 0.801 & 0.061 & 0.579 & 0.546 & 0.546 & 0.144 & 0.757 & 0.738 & 0.798 & 0.023 & 0.280 \\ 
1 & 20 & 20 & 1.0 & 1.3 & 2.0 & 5.0 & 10.0 & 18.0 & 0.1 & 0.5 & 1.0 & 4.00 & 11.00 & 0.50 & 0.09 & 0.121 & 0.815 & 0.109 & 0.579 & 0.546 & 0.546 & 0.231 & 0.757 & 0.738 & 0.798 & 0.041 & 0.339 \\ 
1 & 20 & 20 & 1.0 & 1.4 & 2.0 & 5.0 & 10.0 & 18.0 & 0.1 & 0.5 & 1.0 & 4.00 & 11.00 & 0.50 & 0.09 & 0.174 & 0.827 & 0.177 & 0.579 & 0.546 & 0.546 & 0.346 & 0.757 & 0.738 & 0.798 & 0.072 & 0.414 \\ 
1 & 20 & 20 & 1.0 & 1.5 & 2.0 & 5.0 & 10.0 & 18.0 & 0.1 & 0.5 & 1.0 & 4.00 & 11.00 & 0.50 & 0.09 & 0.227 & 0.850 & 0.281 & 0.579 & 0.546 & 0.546 & 0.451 & 0.757 & 0.738 & 0.798 & 0.112 & 0.493 \\ 
1 & 20 & 20 & 1.0 & 1.6 & 2.0 & 5.0 & 10.0 & 18.0 & 0.1 & 0.5 & 1.0 & 4.00 & 11.00 & 0.50 & 0.09 & 0.301 & 0.875 & 0.386 & 0.579 & 0.546 & 0.546 & 0.600 & 0.757 & 0.738 & 0.798 & 0.152 & 0.505 \\ 
1 & 20 & 20 & 1.0 & 1.7 & 2.0 & 5.0 & 10.0 & 18.0 & 0.1 & 0.5 & 1.0 & 4.00 & 11.00 & 0.50 & 0.09 & 0.350 & 0.893 & 0.507 & 0.579 & 0.546 & 0.546 & 0.710 & 0.757 & 0.738 & 0.798 & 0.189 & 0.540 \\ 
1 & 20 & 20 & 1.0 & 1.8 & 2.0 & 5.0 & 10.0 & 18.0 & 0.1 & 0.5 & 1.0 & 4.00 & 11.00 & 0.50 & 0.09 & 0.387 & 0.919 & 0.646 & 0.579 & 0.546 & 0.546 & 0.818 & 0.757 & 0.738 & 0.798 & 0.236 & 0.610 \\ 
   \hline
\end{tabular}
}
\caption{Simulated power for 4 correlated endpoints in 2-sample design}

\end{table}

%%%%%%%%%%%%%%%%%%%%%%%%%%%%%%%%%%%%%%%%%%%%%%%%%%%%%%%%%%%%
\subsection{Global IUT for both three correlated endpoints when claiming at least noninferiority of standard with two new treatments}
\begin{table}[H]
\centering
\scalebox{0.39}{
\begin{tabular}{rrrrrrrrrrrrrrrr|r|rr|rrrrrr|rrrrrr|r|r}
  \hline
 & n1 & n2 & n3 & ma1 & ma2 & ma3 & mb1 & mb2 & mb3 & mc1 & mc2 & mc3 & sa & sb & sc & rho & IUT & UIT & m1 & m2 & m3 & m4 & m5 & m6 & a1 & a2 & b1 & b2 & c1 & c2 & aiaUIT & RR \\ 
  \hline
	H0 & 20 & 20 & 20 & 1.0 & 1.0 & 1.0 & 2.0 & 2.0 & 2.0 & 10.0 & 10.0 & 10.0 & 1.0 & 4.0 & 11.00 & 0.9& 0.0011 & 0.0474 & 0.0146 & 0.0151 & 0.015 & 0.015 & 0.015 & 0.016 & 0.026 & 0.025 & 0.027 & 0.026 & 0.027 & 0.026 & 0.0004 & 0.364 \\ 
	
	\hline
1 & 20 & 20 & 20 & 1.0 & 1.6 & 1.6 & 2.0 & 5.0 & 5.0 & 10.0 & 18.0 & 18.0 & 1.0 & 4.0 & 11.00 & 0.90 & 0.28 & 0.80 & 0.362 & 0.393 & 0.590 & 0.580 & 0.558 & 0.543 & 0.482 & 0.480 & 0.664 & 0.658 & 0.646 & 0.637 & 0.196 & 0.688 \\ 
1 & 20 & 20 & 20 & 1.0 & 1.7 & 1.7 & 2.0 & 5.0 & 5.0 & 10.0 & 18.0 & 18.0 & 1.0 & 4.0 & 11.00 & 0.90 & 0.36 & 0.80 & 0.501 & 0.511 & 0.574 & 0.592 & 0.550 & 0.569 & 0.594 & 0.599 & 0.652 & 0.676 & 0.623 & 0.645 & 0.273 & 0.756 \\ 
1 & 20 & 20 & 20 & 1.0 & 1.8 & 1.8 & 2.0 & 5.0 & 5.0 & 10.0 & 18.0 & 18.0 & 1.0 & 4.0 & 11.00 & 0.90 & 0.39 & 0.83 & 0.636 & 0.634 & 0.562 & 0.568 & 0.546 & 0.534 & 0.702 & 0.708 & 0.647 & 0.661 & 0.629 & 0.628 & 0.286 & 0.726 \\ 
1 & 20 & 20 & 20 & 1.0 & 1.9 & 1.9 & 2.0 & 5.0 & 5.0 & 10.0 & 18.0 & 18.0 & 1.0 & 4.0 & 11.00 & 0.90 & 0.40 & 0.87 & 0.731 & 0.739 & 0.556 & 0.573 & 0.528 & 0.541 & 0.797 & 0.807 & 0.648 & 0.663 & 0.618 & 0.629 & 0.315 & 0.784 \\ 
  \hline
1 & 20 & 20 & 20 & 1.0 & 1.6 & 1.6 & 2.0 & 5.0 & 5.0 & 10.0 & 19.0 & 19.0 & 1.0 & 4.0 & 11.00 & 0.90 & 0.31 & 0.81 & 0.383 & 0.402 & 0.563 & 0.582 & 0.649 & 0.667 & 0.469 & 0.487 & 0.659 & 0.664 & 0.726 & 0.744 & 0.228 & 0.735 \\ 
1 & 20 & 20 & 20 & 1.0 & 1.7 & 1.7 & 2.0 & 5.0 & 5.0 & 10.0 & 19.0 & 19.0 & 1.0 & 4.0 & 11.00 & 0.90 & 0.37 & 0.83 & 0.500 & 0.495 & 0.561 & 0.573 & 0.628 & 0.637 & 0.585 & 0.595 & 0.662 & 0.656 & 0.719 & 0.720 & 0.284 & 0.768 \\ 
1 & 20 & 20 & 20 & 1.0 & 1.8 & 1.8 & 2.0 & 5.0 & 5.0 & 10.0 & 19.0 & 19.0 & 1.0 & 4.0 & 11.00 & 0.90 & 0.42 & 0.83 & 0.604 & 0.604 & 0.546 & 0.559 & 0.622 & 0.637 & 0.675 & 0.688 & 0.643 & 0.647 & 0.696 & 0.731 & 0.325 & 0.776 \\ 
 1 & 20 & 20 & 20 & 1.0 & 1.9 & 1.9 & 2.0 & 5.0 & 5.0 & 10.0 & 19.0 & 19.0 & 1.0 & 4.0 & 11.00 & 0.90 & 0.45 & 0.88 & 0.720 & 0.737 & 0.575 & 0.584 & 0.636 & 0.641 & 0.788 & 0.803 & 0.647 & 0.666 & 0.713 & 0.728 & 0.373 & 0.822 \\ 
   \hline
1 & 20 & 20 & 20 & 1.0 & 1.6 & 1.6 & 2.0 & 5.0 & 6.0 & 10.0 & 19.0 & 19.0 & 1.0 & 4.0 & 11.00 & 0.90 & 0.30 & 0.92 & 0.414 & 0.390 & 0.595 & 0.838 & 0.667 & 0.659 & 0.491 & 0.484 & 0.683 & 0.886 & 0.747 & 0.739 & 0.230 & 0.764 \\ 
1 & 20 & 20 & 20 & 1.0 & 1.7 & 1.7 & 2.0 & 5.0 & 6.0 & 10.0 & 19.0 & 19.0 & 1.0 & 4.0 & 11.00 & 0.90 & 0.41 & 0.91 & 0.492 & 0.507 & 0.577 & 0.835 & 0.651 & 0.660 & 0.587 & 0.604 & 0.666 & 0.879 & 0.729 & 0.733 & 0.310 & 0.749 \\ 
1 & 20 & 20 & 20 & 1.0 & 1.8 & 1.8 & 2.0 & 5.0 & 6.0 & 10.0 & 19.0 & 19.0 & 1.0 & 4.0 & 11.00 & 0.90 & 0.48 & 0.92 & 0.628 & 0.640 & 0.574 & 0.841 & 0.650 & 0.656 & 0.709 & 0.718 & 0.661 & 0.886 & 0.736 & 0.743 & 0.373 & 0.780 \\ 
1 & 20 & 20 & 20 & 1.0 & 1.9 & 1.9 & 2.0 & 5.0 & 6.0 & 10.0 & 19.0 & 19.0 & 1.0 & 4.0 & 11.00 & 0.90 & 0.51 & 0.93 & 0.746 & 0.743 & 0.588 & 0.832 & 0.664 & 0.641 & 0.809 & 0.817 & 0.678 & 0.885 & 0.729 & 0.713 & 0.417 & 0.822 \\ 
   \hline 
1 & 20 & 20 & 20 & 1.0 & 1.6 & 1.6 & 2.0 & 5.0 & 5.0 & 10.0 & 18.0 & 18.0 & 1.0 & 4.0 & 11.00 & 0.09 & 0.09 & 0.91 & 0.318 & 0.318 & 0.497 & 0.476 & 0.443 & 0.443 & 0.474 & 0.497 & 0.671 & 0.656 & 0.626 & 0.637 & 0.022 & 0.239 \\ 
1 & 20 & 20 & 20 & 1.0 & 1.7 & 1.7 & 2.0 & 5.0 & 5.0 & 10.0 & 18.0 & 18.0 & 1.0 & 4.0 & 11.00 & 0.09 & 0.11 & 0.93 & 0.432 & 0.444 & 0.497 & 0.476 & 0.443 & 0.443 & 0.589 & 0.601 & 0.671 & 0.656 & 0.626 & 0.637 & 0.037 & 0.325 \\ 
1 & 20 & 20 & 20 & 1.0 & 1.8 & 1.8 & 2.0 & 5.0 & 5.0 & 10.0 & 18.0 & 18.0 & 1.0 & 4.0 & 11.00 & 0.09 & 0.14 & 0.94 & 0.534 & 0.545 & 0.497 & 0.476 & 0.443 & 0.443 & 0.688 & 0.713 & 0.671 & 0.656 & 0.626 & 0.637 & 0.051 & 0.364 \\ 
1 & 20 & 20 & 20 & 1.0 & 1.9 & 1.9 & 2.0 & 5.0 & 5.0 & 10.0 & 18.0 & 18.0 & 1.0 & 4.0 & 11.00 & 0.09 & 0.17 & 0.96 & 0.646 & 0.656 & 0.497 & 0.476 & 0.443 & 0.443 & 0.793 & 0.807 & 0.671 & 0.656 & 0.626 & 0.637 & 0.054 & 0.312 \\ 
\hline
1 & 20 & 20 & 20 & 1.0 & 1.6 & 1.6 & 2.0 & 5.0 & 5.0 & 10.0 & 19.0 & 19.0 & 1.0 & 4.0 & 11.00 & 0.09 & 0.11 & 0.94 & 0.318 & 0.318 & 0.497 & 0.476 & 0.556 & 0.562 & 0.474 & 0.497 & 0.671 & 0.656 & 0.724 & 0.747 & 0.026 & 0.232 \\ 
1 & 20 & 20 & 20 & 1.0 & 1.7 & 1.7 & 2.0 & 5.0 & 5.0 & 10.0 & 19.0 & 19.0 & 1.0 & 4.0 & 11.00 & 0.09 & 0.14 & 0.95 & 0.432 & 0.444 & 0.497 & 0.476 & 0.556 & 0.562 & 0.589 & 0.601 & 0.671 & 0.656 & 0.724 & 0.747 & 0.043 & 0.301 \\ 
 1 & 20 & 20 & 20 & 1.0 & 1.8 & 1.8 & 2.0 & 5.0 & 5.0 & 10.0 & 19.0 & 19.0 & 1.0 & 4.0 & 11.00 & 0.09 & 0.18 & 0.96 & 0.534 & 0.545 & 0.497 & 0.476 & 0.556 & 0.562 & 0.688 & 0.713 & 0.671 & 0.656 & 0.724 & 0.747 & 0.064 & 0.354 \\ 
1 & 20 & 20 & 20 & 1.0 & 1.9 & 1.9 & 2.0 & 5.0 & 5.0 & 10.0 & 19.0 & 19.0 & 1.0 & 4.0 & 11.00 & 0.09 & 0.23 & 0.97 & 0.646 & 0.656 & 0.497 & 0.476 & 0.556 & 0.562 & 0.793 & 0.807 & 0.671 & 0.656 & 0.724 & 0.747 & 0.071 & 0.311 \\ 
\hline
1 & 20 & 20 & 20 & 1.0 & 1.6 & 1.6 & 2.0 & 5.0 & 6.0 & 10.0 & 19.0 & 19.0 & 1.0 & 4.0 & 11.00 & 0.09 & 0.14 & 0.97 & 0.318 & 0.318 & 0.497 & 0.766 & 0.556 & 0.562 & 0.474 & 0.497 & 0.671 & 0.894 & 0.724 & 0.747 & 0.035 & 0.259 \\ 
1 & 20 & 20 & 20 & 1.0 & 1.7 & 1.7 & 2.0 & 5.0 & 6.0 & 10.0 & 19.0 & 19.0 & 1.0 & 4.0 & 11.00 & 0.09 & 0.17 & 0.98 & 0.432 & 0.444 & 0.497 & 0.766 & 0.556 & 0.562 & 0.589 & 0.601 & 0.671 & 0.894 & 0.724 & 0.747 & 0.056 & 0.320 \\ 
1 & 20 & 20 & 20 & 1.0 & 1.8 & 1.8 & 2.0 & 5.0 & 6.0 & 10.0 & 19.0 & 19.0 & 1.0 & 4.0 & 11.00 & 0.09 & 0.22 & 0.98 & 0.534 & 0.545 & 0.497 & 0.766 & 0.556 & 0.562 & 0.688 & 0.713 & 0.671 & 0.894 & 0.724 & 0.747 & 0.082 & 0.369 \\ 
1 & 20 & 20 & 20 & 1.0 & 1.9 & 1.9 & 2.0 & 5.0 & 6.0 & 10.0 & 19.0 & 19.0 & 1.0 & 4.0 & 11.00 & 0.09 & 0.28 & 0.99 & 0.646 & 0.656 & 0.497 & 0.766 & 0.556 & 0.562 & 0.793 & 0.807 & 0.671 & 0.894 & 0.724 & 0.747 & 0.095 & 0.342 \\ 
  \hline
1 & 20 & 20 & 20 & 1.0 & 1.7 & 1.7 & 2.0 & 5.0 & 5.0 & 10.0 & 18.0 & 18.0 & 1.0 & 4.0 & 11.00 & 0.70 & 0.28 & 0.84 & 0.449 & 0.488 & 0.524 & 0.540 & 0.500 & 0.509 & 0.588 & 0.628 & 0.662 & 0.670 & 0.620 & 0.643 & 0.177 & 0.623 \\ 
1 & 20 & 20 & 20 & 1.0 & 1.8 & 1.8 & 2.0 & 5.0 & 5.0 & 10.0 & 18.0 & 18.0 & 1.0 & 4.0 & 11.00 & 0.70 & 0.30 & 0.88 & 0.585 & 0.591 & 0.508 & 0.518 & 0.483 & 0.496 & 0.703 & 0.720 & 0.655 & 0.661 & 0.629 & 0.633 & 0.189 & 0.622 \\ 
1 & 20 & 20 & 20 & 1.0 & 1.9 & 1.9 & 2.0 & 5.0 & 5.0 & 10.0 & 18.0 & 18.0 & 1.0 & 4.0 & 11.00 & 0.70 & 0.37 & 0.90 & 0.694 & 0.723 & 0.523 & 0.559 & 0.513 & 0.537 & 0.799 & 0.820 & 0.655 & 0.687 & 0.641 & 0.666 & 0.226 & 0.609 \\ 
\hline
1 & 20 & 20 & 20 & 1.0 & 1.6 & 1.6 & 2.0 & 5.0 & 5.0 & 10.0 & 19.0 & 19.0 & 1.0 & 4.0 & 11.00 & 0.70 & 0.25 & 0.87 & 0.347 & 0.344 & 0.514 & 0.550 & 0.610 & 0.637 & 0.480 & 0.492 & 0.648 & 0.683 & 0.733 & 0.760 & 0.150 & 0.607 \\ 
1 & 20 & 20 & 20 & 1.0 & 1.7 & 1.7 & 2.0 & 5.0 & 5.0 & 10.0 & 19.0 & 19.0 & 1.0 & 4.0 & 11.00 & 0.70 & 0.31 & 0.88 & 0.452 & 0.488 & 0.529 & 0.543 & 0.598 & 0.611 & 0.589 & 0.629 & 0.663 & 0.675 & 0.715 & 0.753 & 0.191 & 0.614 \\ 
1 & 20 & 20 & 20 & 1.0 & 1.8 & 1.8 & 2.0 & 5.0 & 5.0 & 10.0 & 19.0 & 19.0 & 1.0 & 4.0 & 11.00 & 0.70 & 0.36 & 0.90 & 0.587 & 0.593 & 0.520 & 0.526 & 0.606 & 0.598 & 0.709 & 0.728 & 0.664 & 0.659 & 0.728 & 0.733 & 0.217 & 0.610 \\ 
1 & 20 & 20 & 20 & 1.0 & 1.9 & 1.9 & 2.0 & 5.0 & 5.0 & 10.0 & 19.0 & 19.0 & 1.0 & 4.0 & 11.00 & 0.70 & 0.38 & 0.92 & 0.717 & 0.699 & 0.529 & 0.519 & 0.619 & 0.587 & 0.818 & 0.801 & 0.669 & 0.648 & 0.734 & 0.722 & 0.248 & 0.656 \\ 
\hline
1 & 20 & 20 & 20 & 1.0 & 1.6 & 1.6 & 2.0 & 5.0 & 6.0 & 10.0 & 19.0 & 19.0 & 1.0 & 4.0 & 11.00 & 0.70 & 0.25 & 0.90 & 0.330 & 0.331 & 0.512 & 0.797 & 0.593 & 0.596 & 0.460 & 0.477 & 0.656 & 0.867 & 0.726 & 0.730 & 0.140 & 0.562 \\ 
1 & 20 & 20 & 20 & 1.0 & 1.7 & 1.7 & 2.0 & 5.0 & 6.0 & 10.0 & 19.0 & 19.0 & 1.0 & 4.0 & 11.00 & 0.70 & 0.34 & 0.92 & 0.460 & 0.489 & 0.529 & 0.803 & 0.598 & 0.607 & 0.600 & 0.626 & 0.665 & 0.884 & 0.718 & 0.746 & 0.217 & 0.644 \\ 
1 & 20 & 20 & 20 & 1.0 & 1.8 & 1.8 & 2.0 & 5.0 & 6.0 & 10.0 & 19.0 & 19.0 & 1.0 & 4.0 & 11.00 & 0.70 & 0.39 & 0.93 & 0.595 & 0.603 & 0.521 & 0.797 & 0.610 & 0.607 & 0.713 & 0.740 & 0.660 & 0.882 & 0.733 & 0.740 & 0.251 & 0.645 \\ 
1 & 20 & 20 & 20 & 1.0 & 1.9 & 1.9 & 2.0 & 5.0 & 6.0 & 10.0 & 19.0 & 19.0 & 1.0 & 4.0 & 11.00 & 0.70 & 0.45 & 0.94 & 0.686 & 0.717 & 0.525 & 0.799 & 0.638 & 0.619 & 0.802 & 0.814 & 0.655 & 0.883 & 0.759 & 0.751 & 0.293 & 0.650 \\ 
   \hline
\end{tabular}
}
\caption{Simulated power for 3 correlated endpoints in 3-sample design}

\end{table}

%%%%%%%%%%%%%%%%%%%%%%%%%%%%%%%%%%%%%%%%%%
\section{Analysis of an example}
 The availability of real data with multiple treatments and multiple primary endpoints is challenging. Therefore, the data of a dose-finding study to reduce lipids \cite{BakkerArkema1996} are adapted by means of simulation (assumed $\rho=0.8)$. For both endpoints percentage change of triglyceride vs. baseline and  percentage change of cholesterol after 4 weeks as well as the placebo and the three doses 5, 20, 80 mg, the data used here are included in the Appendix and visualized by means of a boxplot:

\begin{figure}[H]
	\centering
		\includegraphics[width=0.3\textwidth]{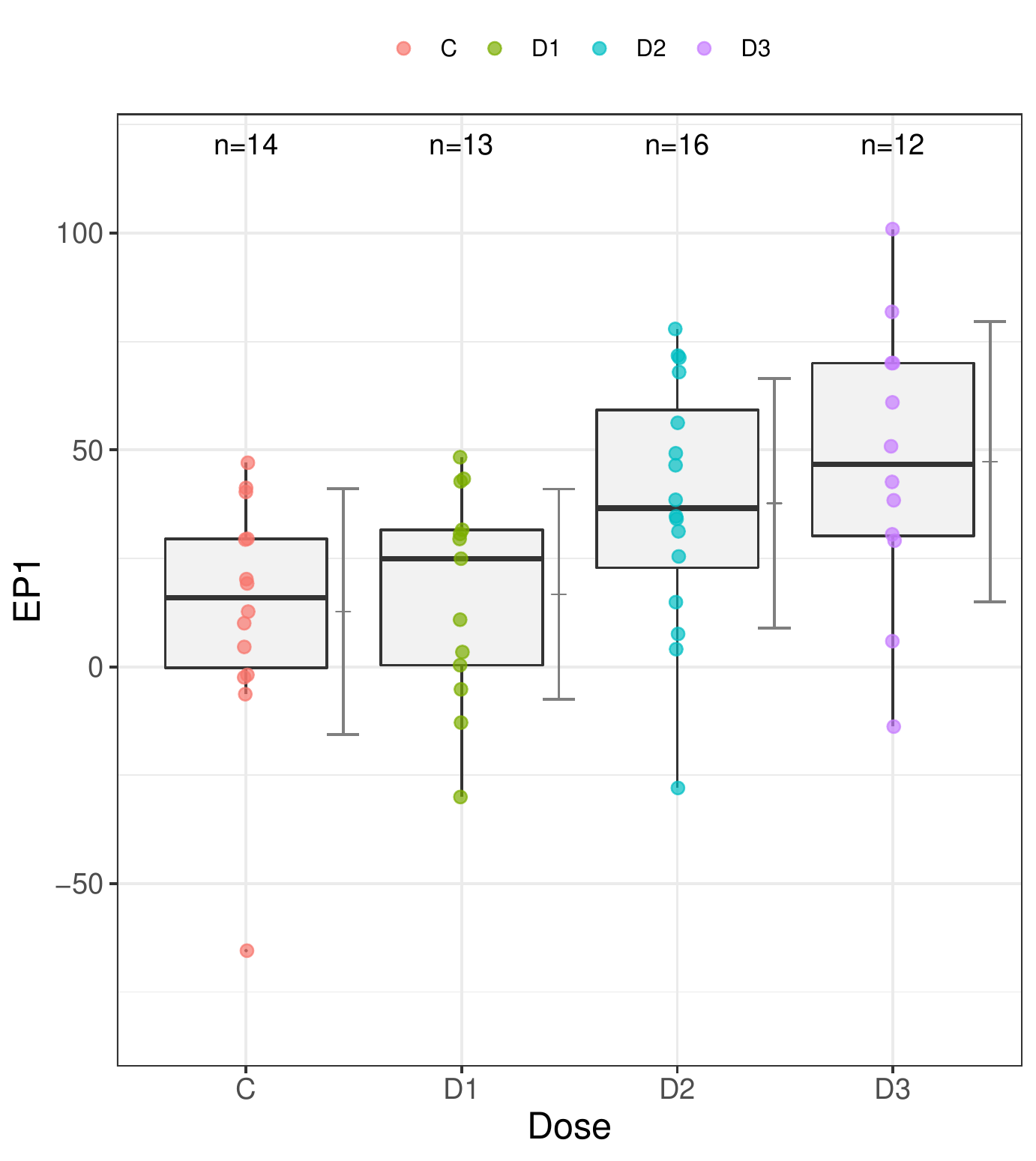}
			\includegraphics[width=0.3\textwidth]{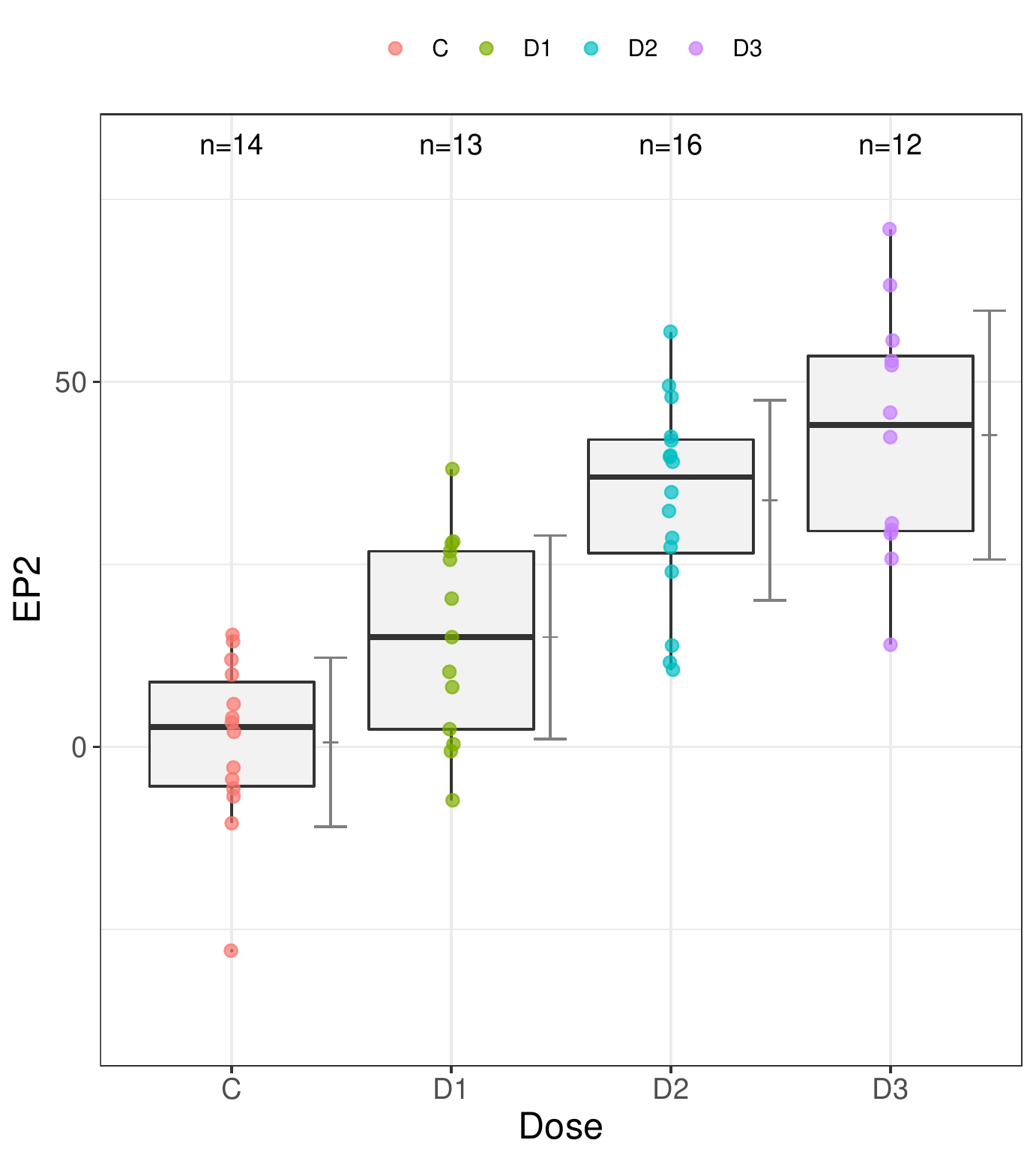}
	\caption{Simulated data for dose finding study with two primary endpoints EP1 and EP2}
	\label{fig:EP1}
\end{figure}

To obtain comparable tests, the multiple endpoint Dunnett test \cite{Hothorn2021aa} and univariate k-sample contrasts dose vs. placebo were used.

\begin{table}[H]
\centering \small
\begin{tabular}{l|l|l}
  \hline
Marginal hypothesis & $p^{aiaIUT}_{j}$ & $p^{IUT}_{j}$ \\ 
  \hline
	EP1, C vs. 5 & 0.681 & 0.341  \\ 
	EP1, C vs. 20 & 0.033 & 0.008  \\ 
	EP1, C vs. 80 & 0.0086 & 0.002  \\ \hline
	EP2, C vs. 5 & 8.47e-03 & 1.96e-03 \\ 
	EP2, C vs. 20 & 5.13e-09 & 5.59e-10  \\ 
	EP2, C vs. 80 & 3.21e-10 & 3.75e-10 \\ \hline
	
   \hline
\end{tabular}
\caption{p-values: marginal vs. aiaUIT-adjusted in the dose finding study}
\end{table}

The (not too realistic) hypothesis that all doses and both endpoints are effective is tested using classical IUT and the new aiaIUT, with adjusted p-values (max(p)-approach) resulting to: $p^{IUT}_{max}$=0.68 $p^{aiaIUT}_{max}$=0.34

Of course, the marginal p-values (of the IUT-test) are in principle smaller than those of the aiaIUT. But both tests lead to the same conclusion: rejection of the global IUT hypothesis.  The decisive advantage of the aiaIUT test can be seen in Figure 2 with the simulated confidence limits. Without further complicated assumption (\cite{Schmidt2015} one can evaluate the changes of the individual contrasts - this is not possible with the classical IUT by definition:

\begin{figure}[htbp]
	\centering
		\includegraphics[width=0.355\textwidth]{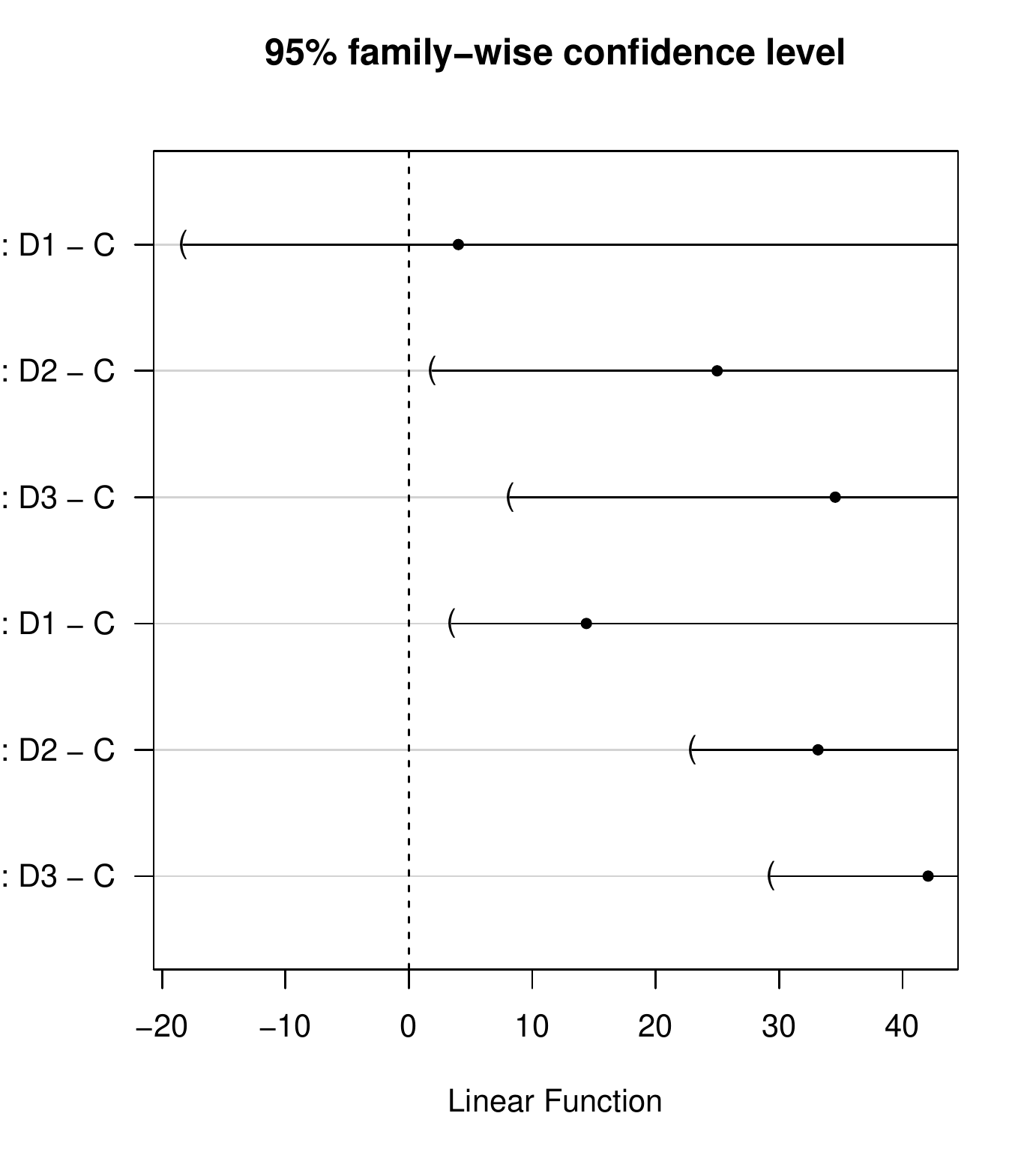}
	\caption{Simultaneous confidence limits for the aiaIUT}
	\label{fig:Aci}
\end{figure}

%%%%%%%%%%%%%%%%%%%%%%%%%%%%%%%%%%%%%%%%%%%%%%%%%%%%%%%
\section{Conclusions}
For the classical intersection-union hypotheses, one can always use the advantages of the complete power UIT as an alternative to the classical IUT, i.e. the additional availability of elementary decisions, special simultaneous confidence intervals, but combined with an inherent power loss.
This power loss becomes smaller (and thus more tolerable given the advantages in a sense of \textit{aesthetics and power considerations in multiple testing} \cite{Hommel2008}) the more elementary tests are included in the hypothesis family and the more highly correlated these tests are, vice versa.
Both test principles are conservative, although with the IUT this becomes extreme as the number of elementary tests increases.
The power is high if all elementary tests are appropriately in the alternative. And it becomes low if at least one is under $H_0$ or close to $H_0$.
\\
Prerequisite is that the multivariate t-distribution with suitable correlation structure for the various max-T tests is numerically available, which is the case in the CRAN packages multcomp and mvtnorm, whereas classical IUT based on simple univariate distributions for the elementary tests.

\footnotesize
\bibliographystyle{plain}
%\bibliography{IUT2021}

\section{Appendix I: R-code for simulation}
\tiny
\begin{verbatim}
#### UIT vs. IUT for 3 correlated endpoints in 3sample design: gaussian
##############################################################################################################
library(tukeytrend); library(multcomp); library(xtable); library(SimComp)
library(CombinePValue); library(robustHD); library(simstudy); library(npmv)
#########################################################################################################
all3DU=function(sims,n1,n2, n3,ma1,ma2,ma3, mb1, mb2,mb3,mc1,mc2,mc3,sa,sb,sc,rho)
{
#sims=50; n1=5; n2=5; n3=5
#ma1=1;ma2=6;ma3=6; mb1=2;mb2=10;mb3=10; mc1=5; mc2=10; mc3=10
#sa=3;sb=7; sc=8; rho=0.8
set.seed(17051949)
  tw = replicate(sims, {
    n<-c(n1,n2,n3); dose<-c("C", "D1", "D2"); Dose = rep(dose, n);
    corMatABC <- matrix(c(1, rho, rho, rho, 1, rho, rho, rho, 1), nrow = 3)
    sdat1 <- genCorData(n1,mu =c(ma1,mb1, mc1), sigma =c(sa,sb,sc), corMatrix = corMatABC)
    sdat2 <- genCorData(n2,mu =c(ma2,mb2, mc2), sigma =c(sa,sb,sc), corMatrix = corMatABC)
    sdat3 <- genCorData(n3,mu =c(ma3,mb3, mc3), sigma =c(sa,sb,sc), corMatrix = corMatABC)
    simdat<-data.frame(Dose = as.factor(Dose), response= rbind(sdat1,sdat2, sdat3))
    colnames(simdat)<-c("Dose", "id", "Y1", "Y2", "Y3")
    fita <- lm(Y1~Dose, data = simdat)        # Endpoint a
    fitb <- lm(Y2~Dose, data = simdat)        # Endpoint b
    fitc <- lm(Y3~Dose, data = simdat)        # Endpoint c
    dfA<-anova(fita)$Df[2]                    # lm df
    multT <-summary(glht(mmm(E1=fita,E2=fitb, E3=fitc),
                         mlf(mcp(Dose ="Dunnett")), alternative="greater", df=dfA)) ### mmm

    MA <-summary(glht(fita,mcp(Dose ="Dunnett"), alternative="greater", df=dfA))
    MB <-summary(glht(fitb,mcp(Dose ="Dunnett"), alternative="greater", df=dfA))
    MC <-summary(glht(fitc,mcp(Dose ="Dunnett"), alternative="greater", df=dfA))

    c(mm=multT$test$pvalues,
      ma=MA$test$pvalues,# univar 1 sided
      mb=MB$test$pvalues,
      mc=MC$test$pvalues)

  })
  erg= as.data.frame(t(tw))
  ####################################################
  estall=length(which(erg$mm1<0.05 & erg$mm2<0.05 & erg$mm3<0.05&
                        erg$mm4<0.05 & erg$mm5<0.05 & erg$mm6<0.05))/sims
  estUI=length(which(erg$mm1<0.05| erg$mm2<0.05 | erg$mm3<0.05|
                       erg$mm4<0.05 | erg$mm5<0.05 | erg$mm6<0.05))/sims
  estm1=length(which(erg$mm1<0.05))/sims
  estm2=length(which(erg$mm2<0.05))/sims
  estm3=length(which(erg$mm3<0.05))/sims
  estm4=length(which(erg$mm4<0.05))/sims
  estm5=length(which(erg$mm5<0.05))/sims
  estm6=length(which(erg$mm6<0.05))/sims
  estma1=length(which(erg$ma1<0.05))/sims
  estma2=length(which(erg$ma2<0.05))/sims
  estmb1=length(which(erg$mb1<0.05))/sims
  estmb2=length(which(erg$mb2<0.05))/sims
  estmc1=length(which(erg$mc1<0.05))/sims
  estmc2=length(which(erg$mc2<0.05))/sims
  estIU=length(which(erg$ma1<0.05 & erg$mb1<0.05 & erg$mc1<0.05
                     &erg$ma2<0.05 & erg$mb2<0.05 & erg$mc2<0.05))/sims
  RR=estall/estIU
  Ergeb=c("n1"=n1, "n2"=n2,"n3"=n3,"ma1"=ma1,"ma2"=ma2, "ma3"=ma3, "mb1"=mb1,
          "mb2"=mb2, "mb3"=mb3, "mc1"=mc1, "mc2"=mc2,"mc3"=mc3,
          "sa"=sa, "sb"=sb, "sc"=sc,  "rho"=rho,
          "IUT"=estIU,  "UIT"=estUI, "m1"=estm1, "m2"=estm2, "m3"=estm3, "m4"=estm4,
          "m5"=estm5, "m6"=estm6,
          "a1"=estma1, "a2"=estma2, "b1"=estmb1, "b2"=estmb2, "c1"=estmc1, "c2"=estmc2,
           "All"=estall, "RR"=RR)
  PX<-xtable(data.frame(t(Ergeb)),
    digits=c(0,0,0,0,1,1,1,1,1,1,1,1,1,1,1,2,6,6,6,6,6,3,3,3,3,3,3,3,3,3,3,6,3))
  print(PX,scalebox = 0.5)

}

all3DU(10000,20,20,20,1,1,1,2,2,2,10,10,10,1,4,11,0.9) # h0

\end{verbatim}

\section{Appendix II: R-code for simulated data example}

\tiny
\begin{verbatim}
library(multcomp)
library(sandwich)
fita <- lm(Y1~Dose, data = simDF)        # Endpoint a
fitb <- lm(Y2~Dose, data = simDF)        # Endpoint b
dfA<-anova(fita)$Df[2]                    # lm df
multT <-summary(glht(mmm(T1=fita,T2=fitb),
                     mlf(mcp(Dose ="Dunnett")), alternative="greater", df=dfA, vcov=sandwich)) ### mmm
plot(glht(mmm(T1=fita,T2=fitb),
                     mlf(mcp(Dose ="Dunnett")), alternative="greater", df=dfA, vcov=sandwich)) ### mmm

MAu <-summary(glht(fita,mcp(Dose ="Dunnett"), alternative="greater", df=dfA, vcov=sandwich), univariate())
MBu <-summary(glht(fitb,mcp(Dose ="Dunnett"), alternative="greater", df=dfA, vcov=sandwich), univariate())
ppp<-cbind(multT$test$pvalue, MAu$test$pvalue, MBu$test$pvalue )

simDF <-
  structure(list(Dose = structure(c(1L, 1L, 1L, 1L, 1L, 1L, 1L,
                                    1L, 1L, 1L, 1L, 1L, 1L, 1L, 2L, 2L, 2L, 2L, 2L, 2L, 2L, 2L, 2L,
                                    2L, 2L, 2L, 2L, 3L, 3L, 3L, 3L, 3L, 3L, 3L, 3L, 3L, 3L, 3L, 3L,
                                    3L, 3L, 3L, 3L, 4L, 4L, 4L, 4L, 4L, 4L, 4L, 4L, 4L, 4L, 4L, 4L
  ), .Label = c("C", "D1", "D2", "D3"), class = "factor"), 
	id = c(1L,2L, 3L, 4L, 5L, 6L, 7L, 8L, 9L, 10L, 11L, 12L, 13L, 14L, 1L,
       2L, 3L, 4L, 5L, 6L, 7L, 8L, 9L, 10L, 11L, 12L, 13L, 1L, 2L, 3L,
       4L, 5L, 6L, 7L, 8L, 9L, 10L, 11L, 12L, 13L, 14L, 15L, 16L, 1L,
       2L, 3L, 4L, 5L, 6L, 7L, 8L, 9L, 10L, 11L, 12L), 
			Y1 = c(-1.8625535108138624,  29.34721445954974, -6.3137678199371745, 19.216991496698075, 12.722266036754805,
             20.212583500058351, 47.051168090719933, 29.513349935229165, 41.257325613773325,
             40.311674165141952, 10.046388081418321, -2.4302074316664015,
             -65.455311923084153, 4.5832349984574563, 31.595255379770506,
             30.657963621804562, 0.37003117275409281, 24.931903758358956,
             -5.1985499511041766, -12.866678239093069, 43.345852071382282,
            -30.019570550674487, 48.327903190301356, 29.475745367759952,
             42.752922177181759, 10.865586982572303, 3.4014910331185852, -27.935908531021326,
             56.258392730398057, 71.745137545619841, 4.0627934257263867, 49.231849215308202,
             77.910315891517342, 67.957178491018084, 38.508391021819591, 7.5764451574917118,
             25.445762179701127, 34.676345429633706, 46.469474749409066, 31.234807114436368,
             34.110312373549867, 14.910507538569316, 71.265170377673641, 81.850999642168176,
             60.967424204246818, 42.627748159404597, 100.92427663976426, 29.08846322726664,
             50.841679855994066, 30.542354253178686, 70.02927830323128, 38.395066333967335,
             5.9093304025316229, -13.78436594698357, 70.033736413459621),
  Y2 = c(-6.777903157821477, 2.0648837823842805, -10.462934365017302,
         14.464374322230817, -2.8380038403019383, 5.8629746208945486,
         3.9939864488029451, 11.949473688213379, 9.9168914824474044,
         15.3395483910324, -4.4397321627653135, -5.6984193139782686,
         -27.91928089342569, 3.3323913097447293, 25.638593326483246,
         27.873863464197768, 2.425097221381467, 8.1926929465386973,
         0.36143222132271902, -7.3254258964661609, 38.054568297829327,
         -0.58317649974408781, 28.139010340561498, 15.031853161767637,
         26.786147976118691, 20.311123560755199, 10.281829889850595,
         10.568567686223858, 42.488334296378078, 56.861102045669298,
         11.562887857471811, 39.913115453262265, 47.929874443514159,
         41.963539317444429, 32.314351990703244, 13.881355717862984,
         23.994996899563738, 34.886671412320652, 39.05634401238963,
         39.718487067123007, 27.372318355449433, 28.629904850643754,
         49.45732631086954, 55.666669246328759, 52.288259549103216,
         29.69764205753204, 70.926629561470435, 25.769983873247657,
         42.426852127674252, 29.209993771785896, 63.247320528419898,
         52.878939449964633, 30.626651541811189, 13.990906423494813,
         45.788680225436714)), 
				class = "data.frame", row.names = c(NA,-55L))


\end{verbatim}
\end{document}